\documentclass[12pt]{article}

\normalbaselines
\RequirePackage{amsmath}
\RequirePackage{amsopn}
\RequirePackage{amsfonts}
\RequirePackage{amsthm}
\RequirePackage{ifthen}

\theoremstyle{definition}

\theoremstyle{remark}

\newtheoremstyle{mytheorem}{0.5cm}{0.2cm}{\slshape}{ }{\bfseries}{.}{ }{}
\theoremstyle{mytheorem}

\newcommand{\EmphFont}{\mdseries\itshape}
\renewcommand{\em}{\EmphFont }

\newcommand{\E}{{\mathbf{E}}}

\usepackage{epsf}
\usepackage{subfigure}

\def\subfigtopskip{0pt}
\def\subfigbottomskip{-2pt}
\def\subfigcapskip{-6pt}

\newcommand{\Rd}{{\mathbf{R}}^d}
\newcommand{\Un}{{\mathsf{Un}}}

\newcommand{\myepsfig}[2]{
  \epsfxsize=#2
  \epsfbox{#1.ps}
  \label{fig:#1}
}

\newcommand{\subfig}[3]{\subfigure[#1]{\myepsfig{#2}{#3cm}}\hspace{-1mm}}

\begin{document}
\bibliographystyle{plain}

\title{Diffusion-limited aggregation with jumps and flights}
\author{I.S.~Molchanov}
\date{}
\maketitle
\begin{abstract}
  The paper suggests a generalisation of the diffusion-limited
  aggregation (DLA) based on using a general stochastic process to
  control particle movements before sticking to a growing cluster.
  This leads to models with variable characteristics that can provide
  a single framework for treating a number of earlier models of
  fractal growth: the DLA, the Eden model and the ballistic
  aggregation. Additionally, a classification of fractal growth models
  is suggested.
\end{abstract}

\section{Introduction}
\label{sec:Introduction}

Stochastic models of fractal growth have inspired a number of studies
and applications in applied sciences. The best known models are the
diffusion-limited aggregation (DLA), the Eden model, and the ballistic
aggregation. These basic models operate according the the following
basic rule.
\begin{quote}
  The initial starting point for the growing cluster is fixed, so that
  the primary cluster consists of a single particle (or pixel for
  simulations on discretised computer screen). At every step another
  pixel (particle) is attached to the cluster according to some rule
  so that the cluster remains connected with respect to some
  neighbouring relationship. This is repeated on every new step until
  the cluster reaches the predetermined size.
\end{quote}
The details depend on the model.  Note that a number of growth models
are discussed in \cite{sta:ost86,vic89}, where further references,
simulated and real pictures and discussions can be found.

\paragraph{The Eden model.} This is a lattice model which was
originally suggested as a model for growth of cell colonies (like
tumours). At any time moment, all neighbours of the active cluster
form the growth zone; a new point to be attached to the cluster is
picked from all these neighbours, where all neighbours have equal
chances to be chosen. Some modifications of the Eden model are
discussed in \cite{mea86}.

\paragraph{DLA.}
This famous model for fractal growth called diffusion-limited
aggregation (DLA) goes back to pioneering papers by Witten and Sander
\cite{wit:san81,wit:san83}.  In this model, every new particle moves
according to Brownian motion starting at infinity (or just far enough
from the growing cluster). When the particle touches the cluster, it
sticks to it irreversibly. This model is called off-lattice DLA in
contrast to lattice DLA where each particle is represented by a pixel
on the grid and moves according to the symmetric random walk on the
grid until it reaches one of the pixels adjacent to the cluster and
then sticks to the cluster. This produces finger-like structure of
fractal type.

The model can be rephrased according to the description of the Eden
model, but where neighbours are not equally likely to be attached to
the cluster.  Instead, the next pixel is chosen among all neighbours
with a distribution proportional to the equilibrium electrostatic
potential on the boundary of the existing cluster, which is the
solution of $\Delta u=0$ where $\Delta=\nabla^2$ is the Laplace
operator.  Therefore, DLA captures the essential features of a typical
dynamic growth process that is related to the Laplace equation.

\paragraph{Ballistic aggregation.} In this model new particles
move along straight lines (ballistic trajectories) until they hit the
cluster or disappear (leave the window of observation). If the mobile
(flying) particle contacts the growing cluster, it sticks at the point
of the first touch.

\paragraph{The aim of the current paper.}
This paper suggests a generalisation of the DLA model which
encompasses a number of known models, including the Eden model, the
standard DLA and the ballistic aggregation. The basic idea is to
replace the Brownian motion in the definition of the standard DLA with
a general stochastic process, possibly having jumps and long-range
dependence.  By varying the characteristics of the stochastic process
it is possible to amalgamate DLA, the Eden model and ballistic
aggregation into a single stochastic growth model. In particular, it
will be shown that the introduced model provides a reasonable tradeoff
between the DLA, ballistic and Eden models and has realisations which
share particular features of the above mentioned models.

We will report several simulation results in two dimensions for
lattice models and moderate numbers of particles in the cluster.
Following the same ideas it is possible to extend simulations to
off-lattice models and increase the size of the cluster by appealing
to efficient algorithms for DLA simulations known from the literature,
see \cite{mea95}.  Finally, we suggest a classification of growth
models, which might be useful in future to unify a number of specific
models known in the literature.

\section{Markov-limited aggregation}
\label{sec:DLA-with-jumps}

The classical DLA model assumes that every new particle moves
according to Brownian motion (or a simple random walk in the lattice
model) starting from infinity (or just sufficiently far from the
cluster).

Brownian motion is just a particular example of a stochastic process
that can be used to induce the particles' movements. Natural
generalisations arise from using other stochastic processes in place
of Brownian motion.  Biased growth was considered in
\cite{ind:lev:gli96}.  In this case a particle performs a biased
random walk where the bias is determined by the particle's position,
so that the corresponding stochastic processes is not spatially
homogeneous.  This leads to DLA-like patterns but with varying fractal
dimension depending on the introduced bias.  Growth in non-Laplacian
fields which appears in the presence of electrostatic field in
addition to the diffusion component was studied in \cite{rob:kna93}.
This is related to the inhomogeneous DLA model considered in
\cite{sel:nit:sta89}.

Let us assume that particles move according to a Markov process
$\xi_t$, $t\geq0$, with values in the $d$-dimensional space $\Rd$ 
%is said to be Markov if the conditional probability of any event
%determined by $\xi_s$ with $s\geq t$ for arbitrary $t$ given $\xi_r$
%for all $r\leq t$ equals the conditional probability of the same event
%given $\xi_t$ only. In other words, this means that the future of the
%process does not depend on its history if the value of the process at
%the present time $t$ is given.  
The Markov process is described by its transition kernel $P(t,x,s,A)$,
which can be identified as the probability that $\xi_s\in A$ given
that $\xi_t=x$.  Loosely speaking, $P(t,x,s,A)$ is the probability of
moving from the state $x$ at time $t$ to the set $A$ at time $s$.
Assume that the process is (time) homogeneous so that $P(t,x,s,A)$
depends on $s-t$ rather than $t$ and $s$ separately. Then the
transition kernel is written as $P_h(x,A)=P(t,x,t+h,A)$.  
%A process is
%spatially homogeneous if $P_h(x,B)=P_h(0,B-x)$, so that $P_h(x,B)$
%depends on $B-x$ rather than on $B$ and $x$ separately.

A Markov process generates two families of operators. One family acts
on bounded functions $f$ as
\begin{displaymath}
  (T_tf)(x)=\int P_t(x,dy)f(y)\,, \quad t\geq0,\; x\in\Rd\,.
\end{displaymath}
Note that $(T_tf)(x)$ can be interpreted as the expected value of
$f(\xi_t)$ given that $\xi_0=x$.  The other family of operators acts
on signed measures $\mu$ with finite total variation (the variation
equals the sum of the absolute values of the positive and negative
components of $\mu$) as
\begin{equation}
  \label{eq:2}
  (T^*_t\mu)(B)=\int P_t(x,B)\mu(dx)\,, \quad t\geq 0\,,
\end{equation}
where $B$ is a measurable subset of $\Rd$. The operators acting on
measures by~\eqref{eq:2} are more important in the sequel, since
$(T^*_t\mu)$ can be interpreted as the distribution of the mass (or
charge if $\mu$ is a signed measure) at time $t$ if the initial
distribution is given by $\mu$ and all elementary masses (or charges)
are being moved along the trajectories of $\xi_t$.

For properties of the Markov process, the behaviour of $T_t$ and
$T^*_t$ as $t\downarrow0$ is crucial. This behaviour is described by
two generating operators which appear as ``derivatives'' of $T_t$ and
$T^*_t$ at $t=0$ and are defined as follows
\begin{align}
  \label{eq:3}
  (Af)(x)&=\lim_{t\downarrow0} t^{-1}((T_tf)(x)-f(x))\,,\\
  \label{eq:4}
  (A^*\mu)(B)&=\lim_{t\downarrow0}t^{-1}((T^*_t\mu)(B)-\mu(B))\,,
\end{align}
provided that the corresponding limits exist. The operators $A$ and
$A^*$ are adjoint, i.e.
\begin{displaymath}
  \langle Af,\mu\rangle=\int (Af)(x)\mu(dx)
  =\int f(x) (A^*\mu)(dx)=\langle f,A^*\mu\rangle
\end{displaymath}
for all $f$ and $\mu$ from the domains of existence of the
corresponding operators. 

It is well known that if $\xi_t$ is the Wiener process (or standard
Brownian motion), then
\begin{displaymath}
  (Af)(x)=\frac{1}{2}(\nabla^2 f)(x)
\end{displaymath}
for twice differentiable $f$ with bounded partial derivatives of the
first and the second order. Here $\nabla^2=\Delta$ is the Laplace
operator, so that 
\begin{displaymath}
  (\nabla^2 f)(x)=\sum_{i=1}^d\frac{\partial^2 f}{\partial x_i^2}(x)\,.
\end{displaymath}
If a signed measure $\mu$ has a sufficiently smooth density $u$, then 
\begin{displaymath}
  (A^*\mu)(dx)=\frac{1}{2}(\nabla^2 u)(x) dx\,.
\end{displaymath}
This fact is widely used in the theory of DLA to relate the properties 
of the growing DLA cluster to the Laplace equation. 

Now define a growth process that is controlled by a general Markov
process $\xi_t$ instead of the Wiener process used to define the
standard DLA. The growth starts with a ball of a small radius $r$
centred at the origin. Let $C$ be the current cluster. At every step a
particle starts at the infinity, which in practice reduces to a random
point $a$ uniformly distributed over the boundary of a sufficiently
large sphere $S$ centred at the origin. The particle moves along the
path of the Markov process $\xi_t$ (starting at $a$) until it either
leaves $S$ or reaches the set $C^r=\{x:\; \rho(x,C)\leq r\}$ (which is
the set of all points at distance no more than $r$ from $C$). In the
first case the particle is killed as soon it leaves $S$, in the second
case, a ball of radius $B_r(x)$ is attached to the cluster, where
$x=\xi_\tau$ is the position of the particle when it first reaches
$C^r$, so that $\tau=\inf\{t\geq0:\; \xi_t\in C^r\}$.

The above description defines an off-lattice aggregation model
determined by the law of the Markov process $\xi_t$. It allows a clear
interpretation in terms of operators associated with the Markov
process (the corresponding explanations for the standard DLA
controlled by the Wiener process can be found in
\cite[Section~18.1]{fal90}).

Let $\mu_t$ denote the distribution of particles at time $t$. Assume
that $\mu_t$ has a density $u(x,t)$, so that $\mu_t(dx)=u(x,t)dx$.
Every particle moves according to the Markov process, so that the
density of particles at point $y$ and time $t+h$ is given by
\begin{equation}
  \label{eq:1}
  (T^*_h\mu_t)(dy)=\mu_{t+h}(dy)=\int P_h(x,dy)\mu_t(dx)\,.
\end{equation}
  Differentiation~\eqref{eq:1} with respect to
$h$ and comparison with~\eqref{eq:4} lead to
\begin{displaymath}
  A^*\mu_t=\frac{d}{dt}\mu_t\,.
\end{displaymath}
In terms of the density $u(x,t)$ the above equation can be written as 
\begin{equation}
  \label{eq:5}
  \frac{\partial u}{\partial t}(x,t)=A^* u(x,t)\,.
\end{equation}
For example, if $\xi_t$ is the standard Brownian motion,
then~\eqref{eq:5} becomes 
\begin{displaymath}
  \frac{\partial u}{\partial t}(x,t)=\frac{1}{2}\nabla^2 u(x,t)\,,
\end{displaymath}
which is the diffusion equation or heat equation. 

If the growth occurs very far from the place where new particles start 
their movement, $u(x)=u(x,t)$ can be considered as independent of $t$, so
that~\eqref{eq:5} turns into
\begin{equation}
  \label{eq:6}
  A^* u(x)=0
\end{equation}
(which becomes $\nabla^2 u=0$ for standard DLA). The boundary
condition is $u=0$ at the boundary of the growing cluster and $u=u_0$
for $\|x\|=R$, where $R$ is the radius of the circle where new
particles are being launched. Solving~\eqref{eq:6} allows us to find
$u$ and then obtain the rate of growth at any point on the cluster's
boundary by computing the derivative of $u$ in the direction of the
normal. Although the direct simulation of the growth process is easier
than numerical solution of~\eqref{eq:6} for the growing cluster,
equation~\eqref{eq:6} provides a useful theoretical interpretation and
justification for the particular choice of the Markov process $\xi_t$,
so that this choice is largely determined by equations that control
the physics of the growth process.

\section{Examples and simulations}
\label{sec:Simulations}

\paragraph{The general setup.}
In simulations on the discrete grid Markov processes are
approximated by their discrete analogues. For example, the Wiener
process becomes a nearest-neighbour symmetrical random walk, and a general
process with independent increments corresponds to a random walk with
varying step length and a possible linear drift. Every new particle
starts on the boundary of a large circle and moves until it either
leaves the circle or becomes a neighbour of the cluster.

When doing simulations on the discrete grid, it is quite typical to
use noise reduction. The essence of noise reduction is that newly
arrived particles are not immediately attached to the growing cluster.
Instead, they are being accumulated so that a new site is attached to
the cluster only when at least $n$ particles have been accumulated
there. If $n=1$, then no noise reduction is present. In the
simulations below we systematically work on the planar square grid and
use the noise reduction parameter $n=4$ (which ensures a moderate
noise reduction while computation time is increased by a factor of 4
only).  The neighbouring relation on the planar square grid is chosen
in such a way that a point (pixel) has 4 neighbours, each sharing one
edge with this pixel.  All simulations below present clusters of size
$52,000$ (unless stated otherwise) simulated on the grid of size
$1024\times1024$ pixels.

\paragraph{The Wiener process and standard DLA.}
Figure~1(a) shows a simulated cluster of 32,000 pixels generated with
$\xi_t$ taken to be standard Brownian motion (or the Wiener process).
The corresponding generating operator is the Laplace operator
$A=(1/2)\nabla^2$, so that the growth corresponds to the models
described by Laplace equation.  

\setcounter{figure}{0}
%\rhead{Figure~\thefigure}
%\vspace*{5cm}
\begin{figure}[htbp]
%  \addtocounter{figure}{1}
  \centering
  \hspace*{-15mm}
  \subfig{The standard DLA (32,000 points)}{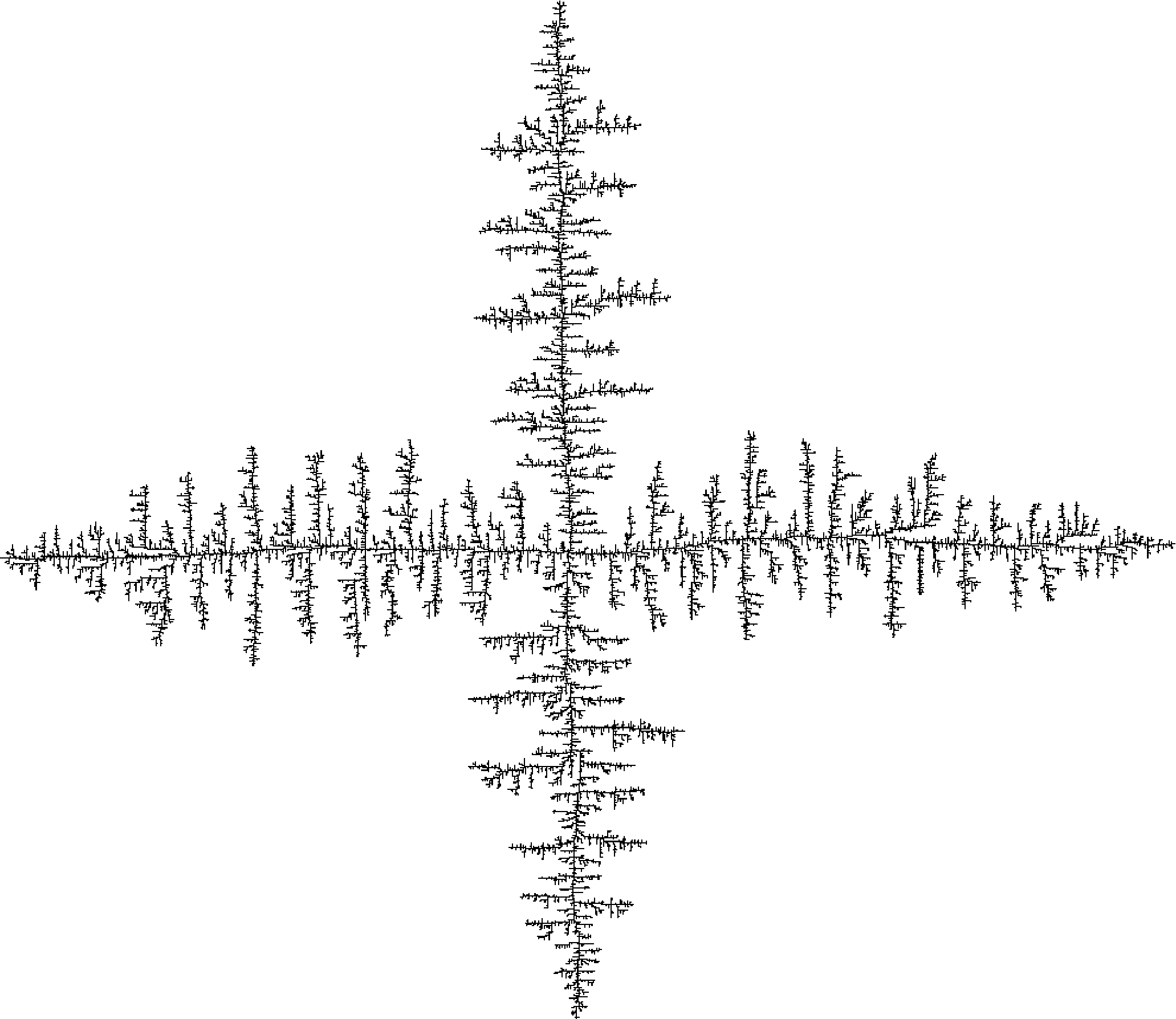}{4}
  \subfig{The Eden model}{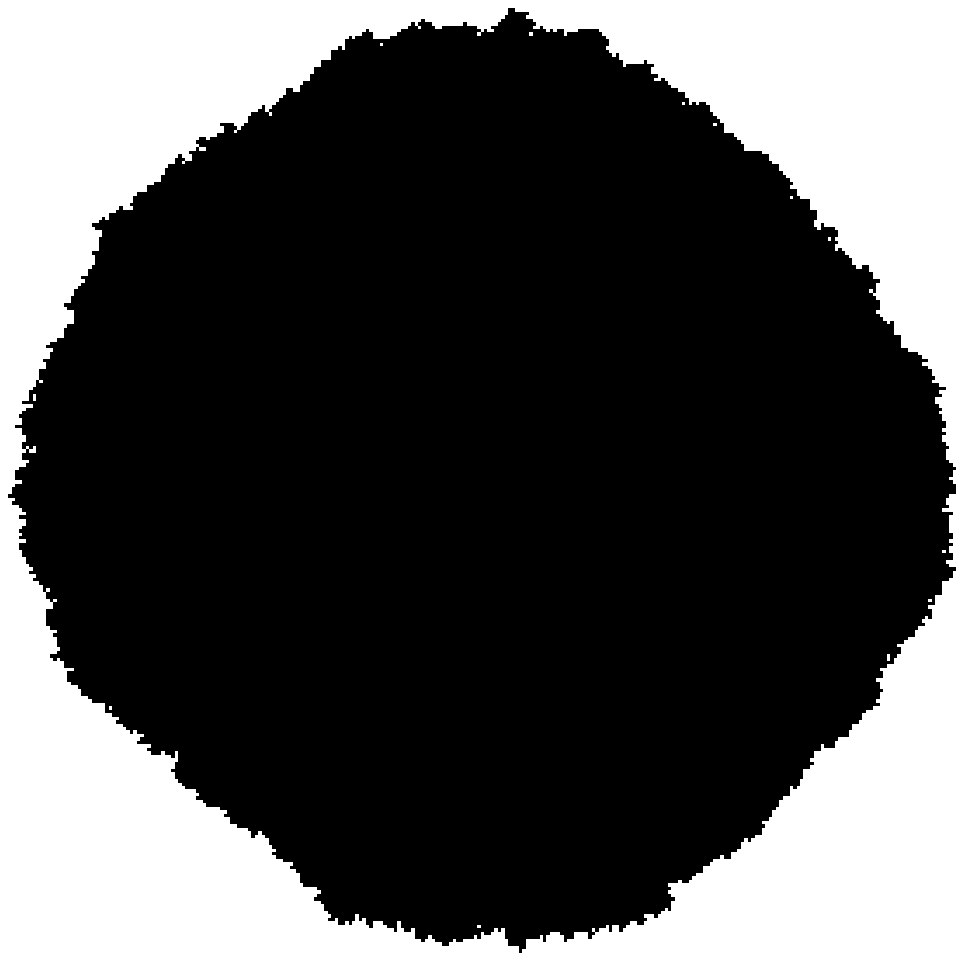}{4}
  \subfig{The circular ballistic model}{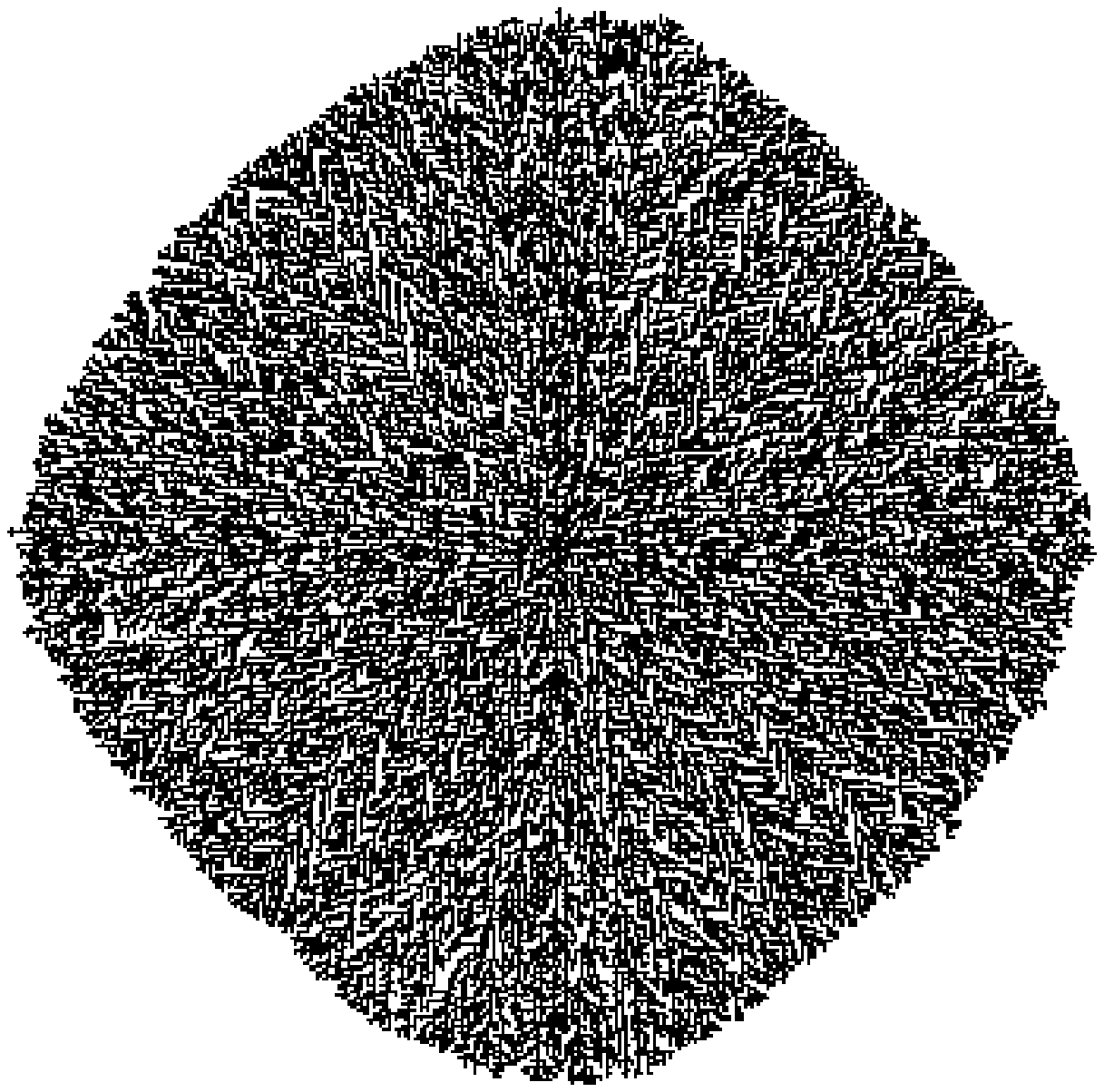}{4}
  \caption{The standard DLA, the Eden model and the
    centre-bound ballistic aggregation.}  
\end{figure}

\paragraph{White noise and the Eden model.}
Assume that $\xi_t$ is a process with independent values (or white
noise). At each step the particle moves to a site chosen completely
random from all points in a disk of a sufficiently large radius.
Figure~1(b) provides a simulated cluster. It is easy to see that the
model so defined is exactly the Eden model. Indeed all pixels from the
growth zone (adjacent to the existing cluster) have equal
probabilities to be chosen, while all other pixels are discarded and
do not influence the evolution. Clearly, from the point of view of
simulations it is more efficient to use a standard way to simulate the
Eden model, in comparison with the white noise simulation where many
points are discarded.  For the white noise case, operators $T_t$ and
$T^*_t$ become identical and independent of $t$, so that the
generating operators are trivial and equation~\eqref{eq:6} becomes the
identity.

\paragraph{Diffusion processes.}
Now assume that $\xi_t$ is a diffusion process. In this case, the
generating operator $A$ is given by a general elliptic operator, so
that~\eqref{eq:6} becomes
\begin{displaymath}
  \frac{1}{2}\sum_{i,j=1}^d b_{ij}(x)
  \frac{\partial^2 u}{\partial x_i\partial x_j}
  -\sum_{i=1}^d a_i(x)\frac{\partial u}{\partial x_i} = 0\,,
\end{displaymath}
where $x=(x_1,\dots,x_d)$, see, e.g., \cite{gih:sko75}.  The matrix
$(b_{ij}(x))_{ij}$ is called the diffusion matrix while $(a_i(x))_i$
is a transfer (drift) vector.  In the spatially homogeneous case
$b_{ij}(x)=b_{ij}$ and $a_i(x)=a_i$ do not depend on $x$.

Note that since diffusion processes are continuous, in the resulting
cluster all points have been visited equally often, which means that
new particles may not end up inside a cluster. This will not be the
case for processes with jumps described below.

\paragraph{Processes with independent increments.}
A stochastic process $\xi_t$ is said to have independent increments if
$\xi_{s_1}-\xi_{s}$ and $\xi_{t_1}-\xi_{t}$ are independent for any
$t\leq t_1\leq s\leq s_1$, so that increments of the process over
disjoint intervals are independent. We assume that the process $\xi_t$
is homogeneous, so that the distribution of $\xi_{t+h}-\xi_t$ does not
depend on $t$. The distribution of a homogeneous process with
independent increments is determined by a transfer vector $(a_i)_i$, a
diffusion matrix $(b_{ij})_{ij}$ and a spectral measure $\Pi$, which
is related to the distribution of jump lengths and the frequencies of
jumps. The generating operator of the process is known for all
stochastically continuous processes with independent increments, see
\cite[p.~344]{gih:sko75}.  This covers processes with jumps and also
with processes having an infinite number of (small) jumps in any
finite time interval. We exclude this latter case and assume that the
spectral measure $\Pi$ is finite. In this case the generating operator
is given by
\begin{equation}
  \label{eq:7}
  (Af)(x)=\sum_{i=1}^d a_i\frac{\partial f}{\partial x_i}
  +\frac{1}{2}\sum_{i,j=1}^d b_{ij}
  \frac{\partial^2 f}{\partial x_i\partial x_j}
  +\int_{\Rd} (f(y+x)-f(x))\Pi(dy)
\end{equation}
and is defined for all bounded twice continuously differentiable
functions $f$ with bounded derivatives of first and second orders. 
In the discrete setup $\Pi(dy)=p_{\mathrm{j}}F(dy)$, where
$p_{\mathrm{j}}$ is the probability of a jump at any given step and
$F(\cdot)$ is the distribution of the jump's length. 

By taking an adjoint operator to $A$ defined in~\eqref{eq:7} we get,
in place of~\eqref{eq:6},
\begin{equation}
  \label{eq:8}
  -\sum_{i=1}^d a_i\frac{\partial u}{\partial x_i}
  +\frac{1}{2}\sum_{i,j=1}^d b_{ij}
  \frac{\partial^2 u}{\partial x_i\partial x_j}
  +\int_{\Rd} (u(x-y)-u(x))\Pi(dy)=0\,.
\end{equation}
It is anticipated that the above integro-differential equation can be
used to model physical processes, where the integral term appears
naturally alongside with the Laplacian or an elliptic operator.

For example, if there is no linear drift and the diffusion matrix is
the unit matrix, then \eqref{eq:8} turns into
\begin{equation}
  \label{eq:9}
  (1-p_{\mathrm{j}})\frac{1}{2}\nabla^2 u
  +p_{\mathrm{j}} \E(u(x-\eta)-u(x))=0\,,
\end{equation}
where $\eta$ is a random vector which describes the jump and $\E$
denotes expectation. Assume that $\eta$ is an isotropic random
vector, so that its distribution is determined by its length
$J=\|\eta\|$. If $J\sim\Un(0,c)$ ($J$ is uniformly distributed over
$[0,c]$), then in the planar case
\begin{displaymath}
  \E(u(x-\eta)-u(x))
  =(2\pi c)^{-1}\int_{B_c(0)} \|y\|^{-1}u(x-y)dy - u(x)\,,
\end{displaymath}
where the integral is taken over the disk of radius $c$ centred at the
origin.  If $p_{\mathrm{j}}=1$, then the model has no diffusion
component and~\eqref{eq:9} becomes
\begin{equation}
  \label{eq:10}
  \E u(x-\eta)=u(x)\,.
\end{equation}
If $J=c$ is a constant, then~\eqref{eq:10} turns into
\begin{equation}
  \label{eq:11}
  \frac{1}{2\pi c} \int_{\|y\|=c} u(x-y)dy =  u(x)\,,
\end{equation}
so that the value of $u$ at any point is equal to the mean value of
the function on the circle of radius $c$ centred at this point. 
Note that in contrast to the definition of a general harmonic
function, \eqref{eq:11} is not required for all $c>0$ but just for a
fixed value of $c$. 

\setlength{\topmargin}{-3cm}
%\vspace*{-4cm}
\setcounter{subfigure}{0}
\begin{figure}[htbp]
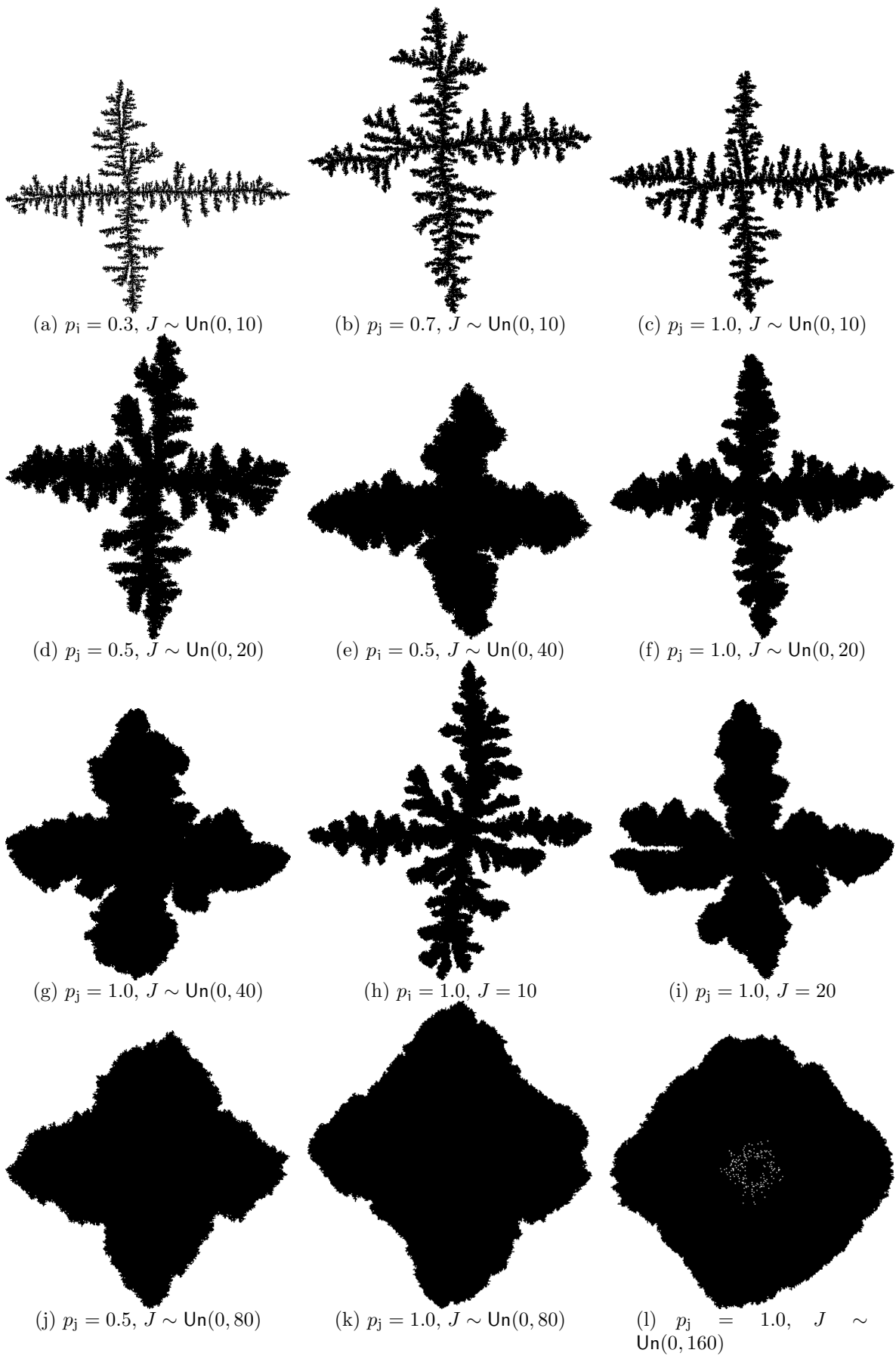

%  \addtocounter{figure}{1}
  \centering
  \hspace*{-15mm}
  \subfig{$p_{\mathrm{j}}=0.3$, $J\sim\Un(0,10)$}{Fig2/w03j}{5}
  \subfig{$p_{\mathrm{j}}=0.7$, $J\sim\Un(0,10)$}{Fig2/w07j}{5}
  \subfig{$p_{\mathrm{j}}=1.0$, $J\sim\Un(0,10)$}{Fig2/w10j}{5}

  \hspace*{-15mm}
  \subfig{$p_{\mathrm{j}}=0.5$, $J\sim\Un(0,20)$}{Fig2/w05j20}{5}
  \subfig{$p_{\mathrm{j}}=0.5$, $J\sim\Un(0,40)$}{Fig2/w05j40}{5}
  \subfig{$p_{\mathrm{j}}=1.0$, $J\sim\Un(0,20)$}{Fig2/w10j20}{5}

  \hspace*{-15mm}
  \subfig{$p_{\mathrm{j}}=1.0$, $J\sim\Un(0,40)$}{Fig2/w10j40}{5}
  \subfig{$p_{\mathrm{j}}=1.0$, $J=10$}{Fig2/w10j10f}{5}
  \subfig{$p_{\mathrm{j}}=1.0$, $J=20$}{Fig2/w10j20f}{5}

  \hspace*{-15mm}
  \subfig{$p_{\mathrm{j}}=0.5$, $J\sim\Un(0,80)$}{Fig2/w05j80}{5}
  \subfig{$p_{\mathrm{j}}=1.0$, $J\sim\Un(0,80)$}{Fig2/w10j80}{5}
  \subfig{$p_{\mathrm{j}}=1.0$, $J\sim\Un(0,160)$}{Fig2/w10j160}{5}
  \caption{Clusters obtained for Markov processes with jumps.}
\end{figure}

Figure~2 shows a number of simulated patterns obtained for the model
which combines the Brownian motion component with jumps. It is easy to
observe that the presence of jumps makes the patterns thicker while
still retaining the fine structure of their boundaries (see
enlargements in Figure~7). Larger jumps make clusters more compact, so
that in the extreme case we can grow patterns which are similar to
those from the Eden model.

The similar thickening effect is achieved within standard DLA by
allowing particles to continue travelling after they become neighbours
of the cluster for the first time, so that the particle can either
stick to the cluster with some probability $p$ or continue travelling
with probability $1-p$, see \cite{wit:san83}.  Then the particles can
penetrate the cluster yielding thickening effect. However, the current
setup naturally establishes relationships of this effect to some
integro-differential equations.

Note that $J\sim\Un(0,c)$ results in a thicker cluster than $J=c/2$
(i.e. with $J$ being the expected value of the former). It should be
noted that this thickening effect is entirely different from the
effect that would be achieved if we enlarge the standard DLA model by
a ball of some radius. The latter results in smoothing out the
boundaries, while the clusters shown in Figure~2 have rough
boundaries, as is clear from Figure~7.

\setcounter{subfigure}{0}
\begin{figure}[p]
%  \addtocounter{figure}{1}
  \centering
  \hspace*{-15mm}
  \subfig{$J=1$ (size 52,400)}{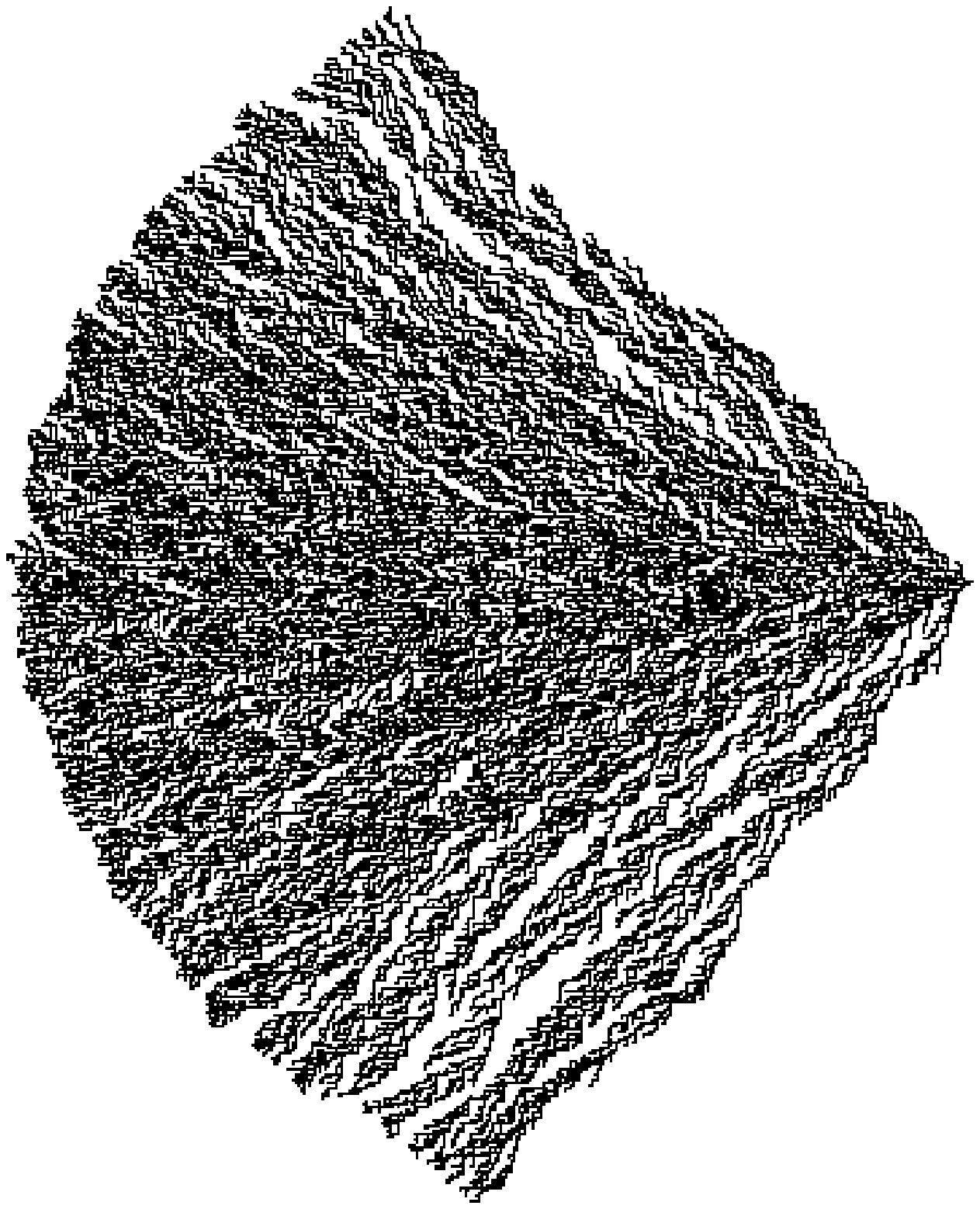}{4.5}\hspace{1cm}
  \subfig{$J=2$}{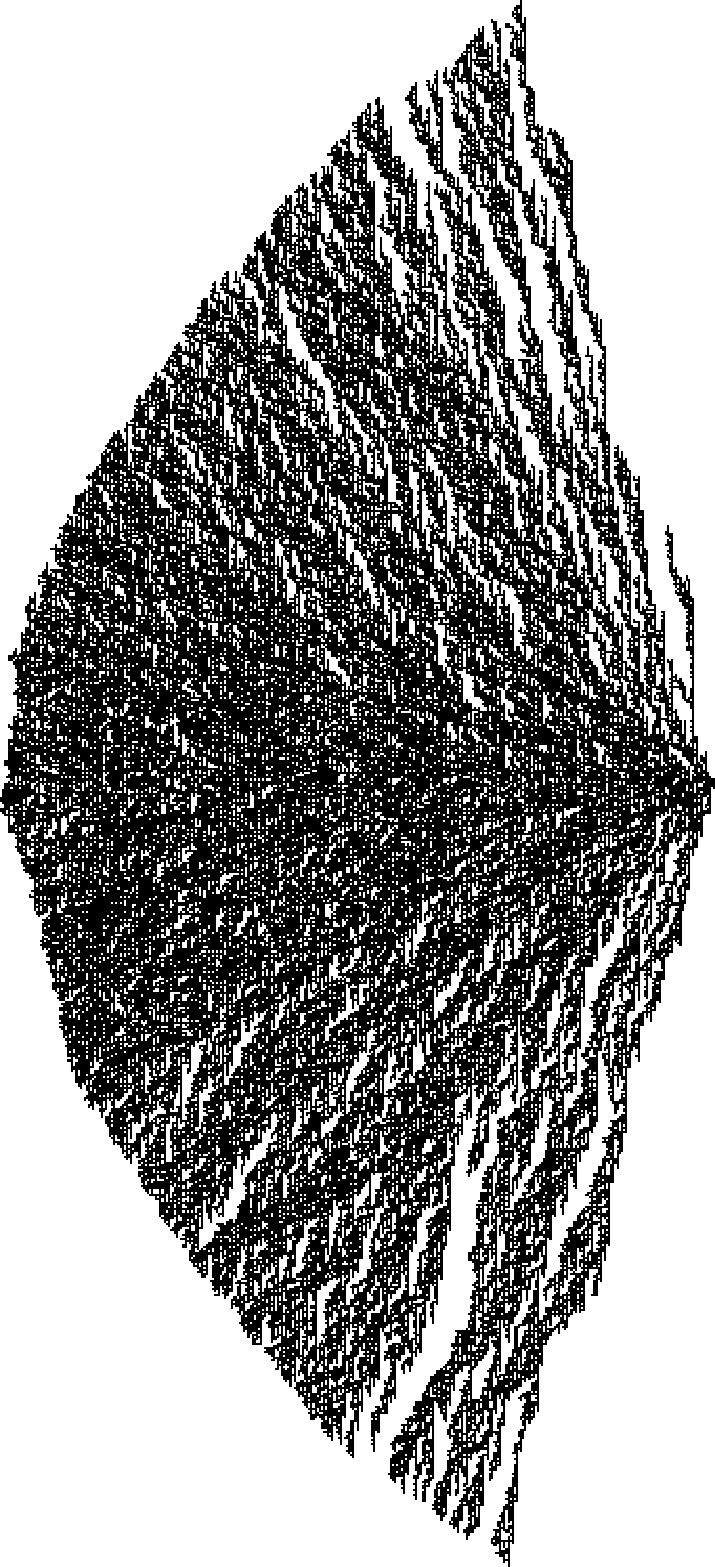}{3}\hspace{1cm}
  \subfig{$J=20$}{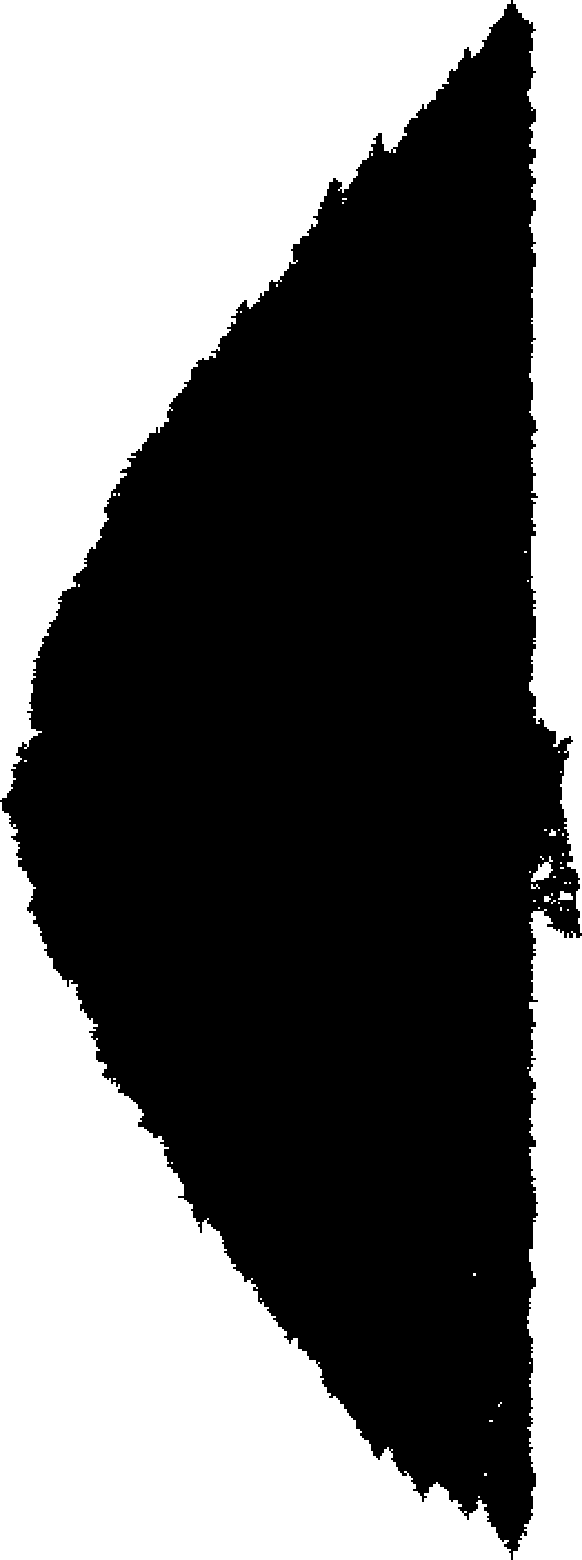}{2.5}

  \hspace*{-5mm}
  \subfig{$J\sim\Un(0,10)$}{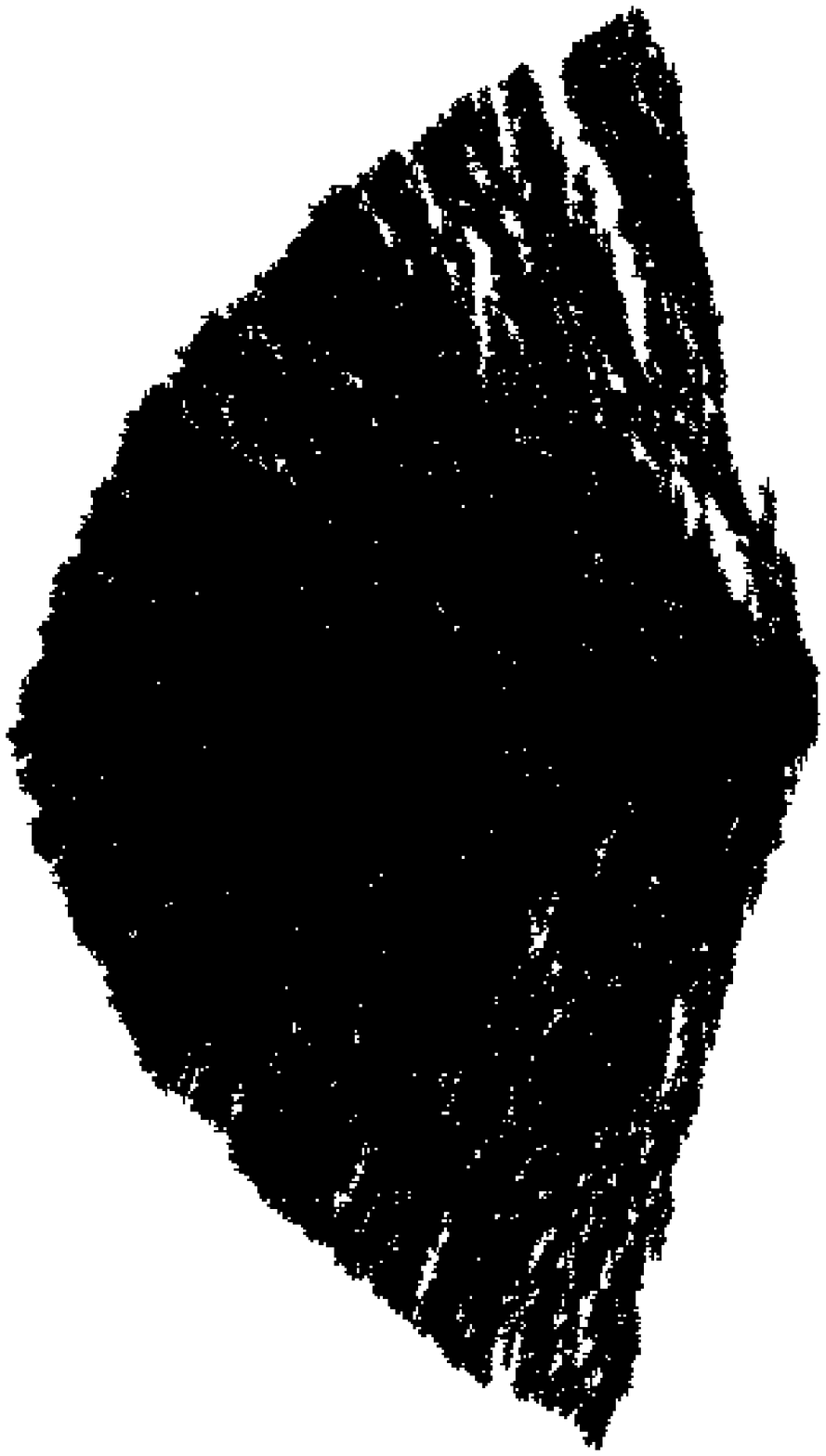}{4}\hspace{1cm}
  \subfig{$J=50$}{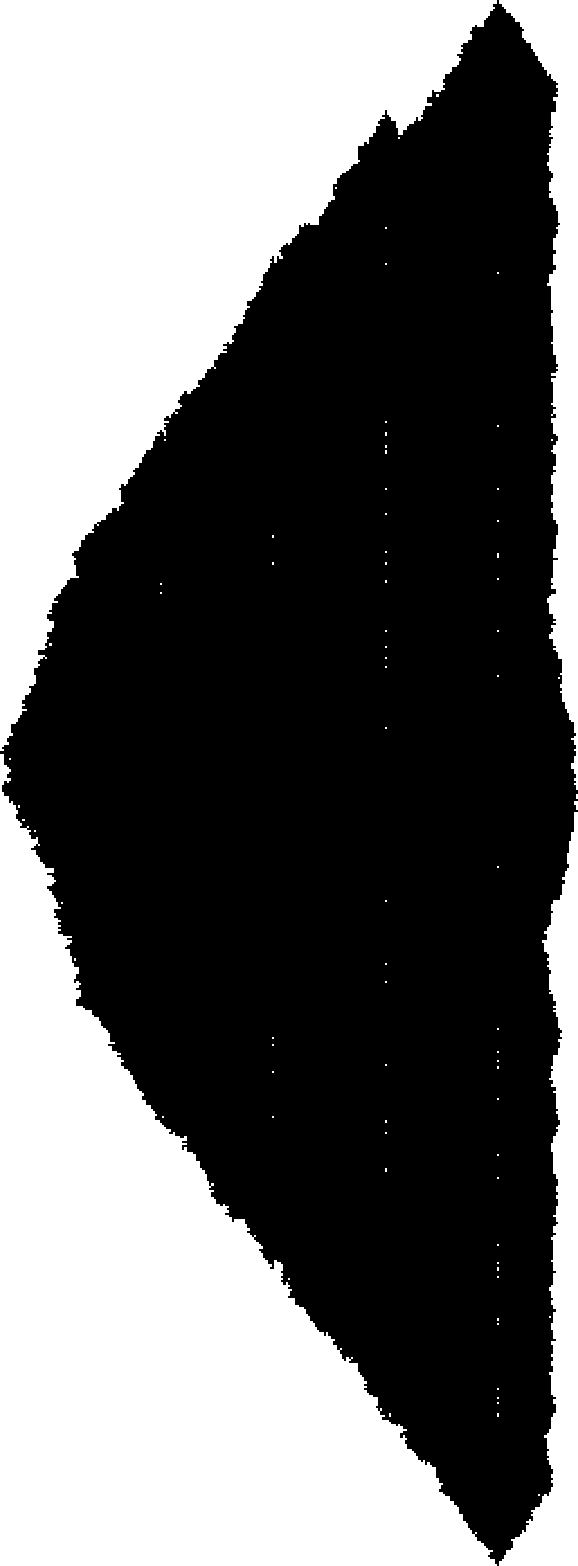}{2.7}\hspace{1cm}
  \subfig{$J=100$}{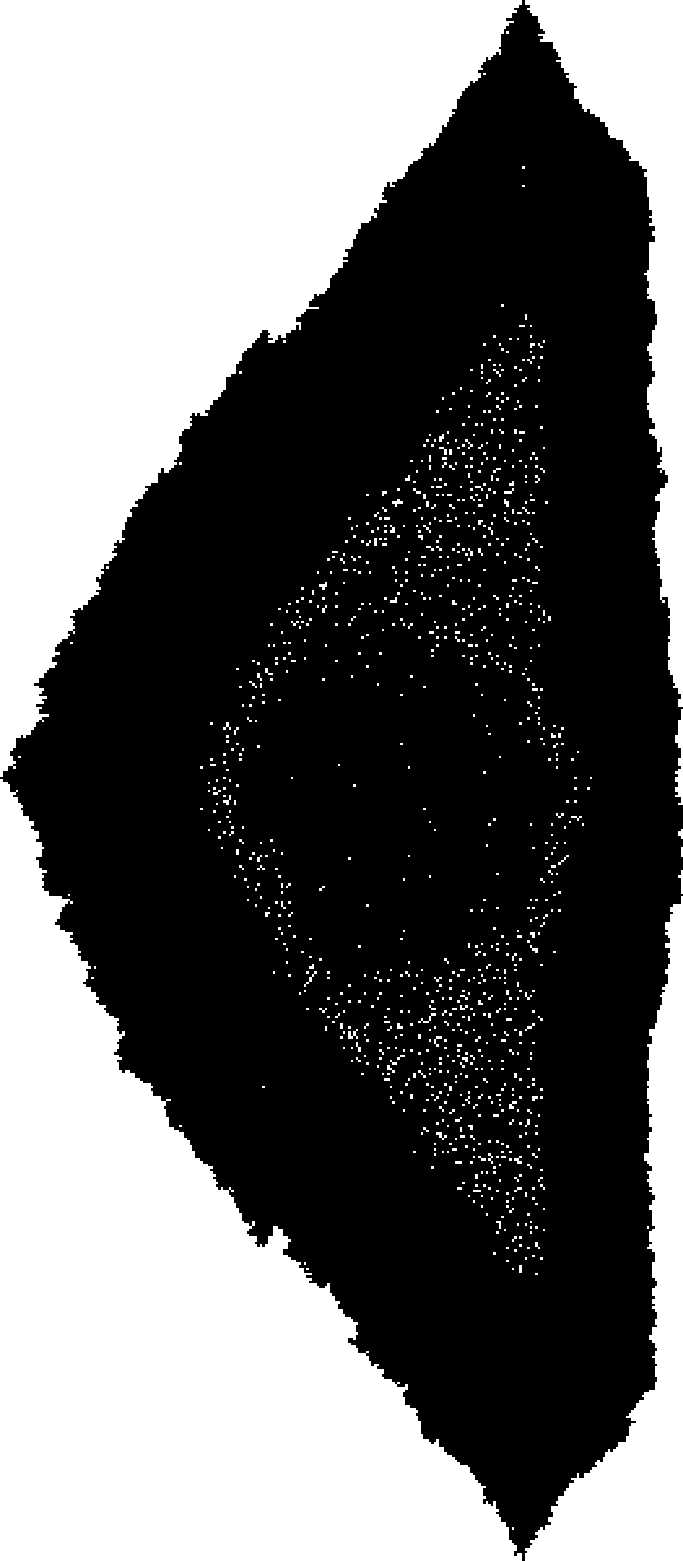}{2.7}
  \caption{Clusters obtained for the Markov model with
    $p_{\mathrm{j}}=1$ and jumps of fixed direction (horizontal to the
    right). The length of jumps $J$ is specified in every particular
    case.  The size of clusters is 104,800 pixels unless specified
    otherwise.}
\end{figure}

Figure~3 shows clusters obtained for pure jump processes with jumps of 
a fixed  direction. Assume that the jump $\eta$ has only a
non-negative $x$ component, so that $\eta=(J,0)$ with $J>0$. All
particles start from points uniformly distributed near the left border 
of the window and move to the right until they leave the window or
stick to the cluster. It is not possible to let all particles start
exactly on the left-hand side of the window, since this may lead to
unwanted periodicity effects which are not compliant with the fact that 
ideally the particles are coming from infinity. To simulate coming 
from infinity, all particles start at points uniformly distributed
within a rectangle located along the left border of the window and
having width equal to the maximum possible value of $J$. 

In particular if $J=c$ is constant, then~\eqref{eq:10} turns into
\begin{displaymath}
  u(x-\eta)=u(x)\,.
\end{displaymath}
Extending $u$ from its value on the points where particles start, we
see that all points that can be reached from the left by moving
(jumping) particles are equally likely to be attached to the cluster.
In particular, if $J=1$, then all neighbours to the cluster which are
visible from the left can be attached to the cluster with equal
probabilities. For larger $J$ inner pixels can also be attached within
a distance of at most $J$ from the left border of the growing cluster.
This leads to the growth similar to that of the Eden model, but with a
clearly expressed directionality.

It is interesting to note that the presence of jumps means that new
particles can end up inside the cluster with their final jump. As a
result of these final jumps they can end up in one of the already
occupied sites, increasing the concentration of particles deposited at
this particular site but without changing the shape of the cluster.
Therefore, the defined growth model with jumps also provides a
function that shows the number of particles accumulated at 
particular sites. 
%Figure~4 shows several renormalised grey-scale
%images where darker values correspond to high particle densities,
%along with the corresponding histograms of the numbers of points
%accumulated at pixels from some clusters shown in Figure~2.

%\setcounter{subfigure}{0}
%\begin{figure}[htbp]
%%  \addtocounter{figure}{1}
%  \centering
%  \renewcommand{\thesubfigure}{}
%  \subff{}{Fig4/w10j20G} 
%  \subff{}{Fig4/w10j40G} 
%  \subff{}{Fig4/w10j80G}

%  \setcounter{subfigure}{0}
%  \renewcommand{\thesubfigure}{(\alph{subfigure})}
%  \subff{$J\sim\Un(0,20)$}{Fig4/w10j20H} 
%  \subff{$J\sim\Un(0,40)$}{Fig4/w10j40H} 
%  \subff{$J\sim\Un(0,80)$}{Fig4/w10j80H}
%  \caption{Grey-scale images representing densities of the particles
%    for Markov models with jumps. The images are truncated to fit
%    window of size $400\times400$. All models have $p_{\mathrm{j}}=1$.
%    The second row presents the histograms of non-zero values from the
%    corresponding images from the first row. The values of the
%    function correspond to the numbers of sites visited particular
%    number of times.}
%\end{figure}

\section{A general model with flights and jumps}
\label{sec:general-model-with}

\paragraph{Non-Markov particle movement.}
The stochastic process $\xi_t$ that determines particles' movements
can be rather general and not necessarily Markov. One example worth
special consideration describes ballistic trajectories of particles.
We say that $\xi_t=\beta (t-t_0)+a$ determines a flight which starts
at point $a$ at time $t_0$ and has a velocity vector $\beta$.  The
basic model considered in this section deals with a stochastic process
that has several components: a diffusion component, pure jumps, and a
ballistic (or flight) component.  For simulations on the discrete
grid, we fix $p_{\mathrm{j}}$, the probability of a jump, and
$p_{\mathrm{b}}$, the probability of a flight. This means that at any
time $\xi_t$ performs a nearest neighbour random walk (or diffusion)
with probability $1-p_{\mathrm{j}}-p_{\mathrm{b}}$, otherwise it jumps
at vector $\eta$ (which is usually isotropic and has length $J$) or
flies with probability $p_{\mathrm{b}}$, with the flight direction
determined by a unit vector $\beta$ along the straight line at
distance $L$.

There are two basic options for the distribution of $\beta$ which will
be explored later on. First, $\beta$ can be an isotropic random
vector. In the second case, $\beta$ depends on the position of the
particle when it starts flying and points to the centre of the
window (or towards the initial point of the cluster). In the first
case we will speak about isotropic flights and in the second case
about centre-bound flights. Clearly, mixtures of these two cases are
also possible, but we will not discuss them here.

Unfortunately the process with flights is not Markov and so the model
does not allow direct interpretation in terms of the generating
operators. 

\paragraph{Circular ballistic model.}
Figure~1c provides an example of a circular ballistic model. In this
model every new particle starts from a point uniformly distributed on
a large circle. This point moves straight towards the centre of the
cluster until it hits the cluster. In this case $p_{\mathrm{b}}=1$,
the distance $L$ is sufficiently large (at least as large as the
radius of the circle) and the random vector $\beta$ is a deterministic
vector which depends on the position of the particle when it starts
flying and is always directed towards the centre of the circle (as in
the centre-bound case).

\paragraph{DLA with flights.}
Figure~4 shows several clusters obtained for mixtures of processes
with independent increments and possible flights. The directions of the 
flights are isotropic for all simulations shown in Figure~4. 

\setcounter{subfigure}{0}
\begin{figure}[htbp]
%  \addtocounter{figure}{1}
  \centering
  \hspace*{-15mm}
  \subfig{$p_{\mathrm{j}}=0$, $p_{\mathrm{b}}=0.3$,
    $L=10$}{Fig5/bw00j03b}{5.5}
  \subfig{$p_{\mathrm{j}}=0$, $p_{\mathrm{b}}=1.0$,
    $L=10$}{Fig5/bw00j10b}{5.5}
  \subfig{$p_{\mathrm{j}}=0$, $p_{\mathrm{b}}=1.0$,
    $L=40$}{Fig5/bw00j10b40}{5.5}

  \hspace*{4mm}
  \subfig{$p_{\mathrm{j}}=0.3$, $J\sim\Un(0,10)$,\protect\newline
    $p_{\mathrm{b}}=0.7$, $L=10$}{Fig5/bw03j07b}{4.5}
  \subfig{$p_{\mathrm{j}}=0.3$, $J\sim\Un(0,40)$,\protect\newline
    $p_{\mathrm{b}}=0.3$, $L=10$}{Fig5/bw03j40u03b}{4.5}
  \subfig{$p_{\mathrm{j}}=0$, $p_{\mathrm{b}}=1.0$,
    $L=80$}{Fig5/bw00j10b80}{4.5} 

  \hspace*{4mm}
  \subfig{$p_{\mathrm{j}}=0.3$, $J\sim\Un(0,10)$,\protect\newline
   $p_{\mathrm{b}}=0.3$, $L=20$}{Fig5/bw03j03b20}{4.5}
  \subfig{$p_{\mathrm{j}}=0.3$, $J\sim\Un(0,10)$,\protect\newline
    $p_{\mathrm{b}}=0.3$, $L=40$}{Fig5/bw03j03b40}{4.5}
  \subfig{$p_{\mathrm{j}}=0.5$, $J=10$,\protect\newline
    $p_{\mathrm{b}}=0.5$,
    $L=5$}{Fig5/bw05j10f05b05}{4.5}
  \caption{Ballistic aggregation with possible jumps and diffusion
    component.}
\end{figure}

It is easy to see that the increase in flight length $L$ results in
models with fine structure (Figure~4c) still adhering to the typical
structure of the standard DLA model. Jumps make patterns thicker in the 
same manner as they do for the model without flights. 

\paragraph{Centre-bound flights.}
A number of examples of DLA with centre-bound flights are shown in
Figure~5. At every step with probability $p_{\mathrm{b}}$ the particle 
flies towards the origin a distance length $L$. 
Further simulations (not shown in Figure~5) show that if $L$ or
$p_{\mathrm{b}}$ increase, then the realisations of the model look
very similar to the circular ballistic model shown in Figure~1c.

\setcounter{subfigure}{0}
\begin{figure}[htbp]
%  \addtocounter{figure}{1}
  \centering
  \hspace*{-15mm}
  \subfig{$p_{\mathrm{j}}=0$, $p_{\mathrm{b}}=0.1$, $L=5$}{Fig6/z00j01b05}{4}
  \subfig{$p_{\mathrm{j}}=0$, $p_{\mathrm{b}}=0.1$, $L=10$}{Fig6/z00j01b}{4}
  \subfig{$p_{\mathrm{j}}=0$, $p_{\mathrm{b}}=0.05$, $L=10$}{Fig6/z00j005b}{4}

  \hspace*{-15mm}
  \subfig{$p_{\mathrm{j}}=0$, $p_{\mathrm{b}}=0.01$, $L=10$}{Fig6/z00j001b}{4}
  \subfig{$p_{\mathrm{j}}=0$, $p_{\mathrm{b}}=0.3$,
    $L=10$}{Fig6/z00j03b}{4}
  \subfig{$p_{\mathrm{j}}=0.99$, $J\sim\Un(0,10)$,\protect\newline
    $p_{\mathrm{b}}=0.01$, $L=10$}{Fig6/z099j001b}{4}

  \hspace*{-15mm}
  \subfig{$p_{\mathrm{j}}=0.9$, $J\sim\Un(0,10)$,\protect\newline
    $p_{\mathrm{b}}=0.1$, $L=5$}{Fig6/z09j01b05}{4}
  \subfig{$p_{\mathrm{j}}=0.9$, $J\sim\Un(0,10)$,\protect\newline
    $p_{\mathrm{b}}=0.1$, $L=10$}{Fig6/z09j01b}{4}
  \subfig{$p_{\mathrm{j}}=0.9$, $J\sim\Un(0,80)$,\protect\newline
    $p_{\mathrm{b}}=0.1$, $L=80$}{Fig6/z09j80u01b80}{4}
  \caption{Centre-bound ballistic aggregation with possible jumps
    and diffusion component. }
\end{figure}

\paragraph{Shooting models.} 
It is possible to define the following interesting variation of the
DLA model with flights and jumps. Assume that all ballistic pieces of
the trajectories are discarded if the particle does not hit the
cluster while flying. In other words, if at any time the particle does
not touch the cluster while flying, then it returns to the start of
the flight and the last flight is discarded. This model can be
naturally called the shooting model, assuming that a rifleman moves
along a Markov process trying to shoot the cluster. If the rifleman
hits the cluster with a shot then a new particle is attached to the
cluster at the place of the first hit. If the rifleman does not hit
the cluster, then he continues travelling, so that at any step he may
attempt another shot with probability $p_{\mathrm{b}}$.
Figure~6 shows several examples of the clusters obtained for the
shooting model. Some of them have isotropic directions of shots
(flights), while others have centre-bound shots (flights). 

\setcounter{subfigure}{0}
\begin{figure}[htpp]
%  \addtocounter{figure}{1}
  \centering
  \subfig{$p_{\mathrm{j}}=0$, $p_{\mathrm{b}}=0.1$, \protect\newline
    $L=10$ (isotropic), \protect\newline cluster size 23,000}{Fig7/J00j01b}{5}
  \subfig{$p_{\mathrm{j}}=0.9$, $J\sim\Un(0,10)$, \protect\newline
    $p_{\mathrm{b}}=0.1$, $L=10$ (isotropic)}{Fig7/J09j01b}{5}

  \subfig{$p_{\mathrm{j}}=0$, $p_{\mathrm{b}}=0.1$, \protect\newline
    $L=10$ (centre-bound),\protect\newline cluster size
    34,000}{Fig7/JB00j01b}{5}
  \subfig{$p_{\mathrm{j}}=0.9$, $J\sim\Un(0,10)$, \protect\newline
    $p_{\mathrm{b}}=0.1$, $L=10$ (centre-bound)}{Fig7/JB09j01b}{5}
  \caption{The shooting model with isotropic flights (a),(b) and
    centre-bound flights (c),(d).}
\end{figure}

It is also possible to work out variations of the shooting model.  For
example, the jumps may also be discarded unless the final point of the
jump is a neighbour to the cluster. This would correspond to a man
throwing grenades and/or shooting from a gun.  

\def\subfigtopskip{0pt}
\def\subfigbottomskip{2pt}
\def\subfigcapskip{6pt}
\newcommand{\subfe}[2]{\subfigure[#1]{\myepsfig{#2}{3cm}}\hspace{-1mm}}
\setcounter{subfigure}{0}
\renewcommand{\thesubfigure}{}
\begin{figure}[htbp]
%  \addtocounter{figure}{1}
  \centering
  \hspace*{-15mm}
  \subfe{Fig1a}{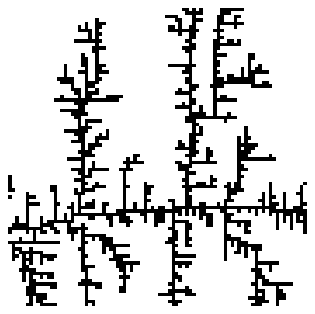}
  \subfe{Fig1b}{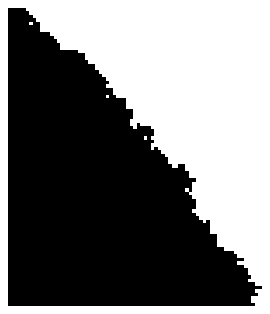}
  \subfe{Fig1c}{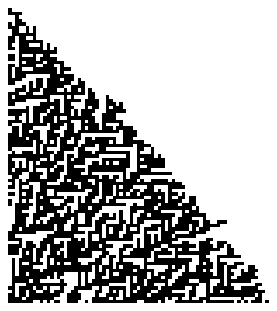}
  \subfe{Fig2a}{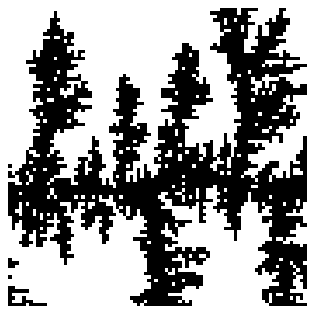}
  \subfe{Fig2c}{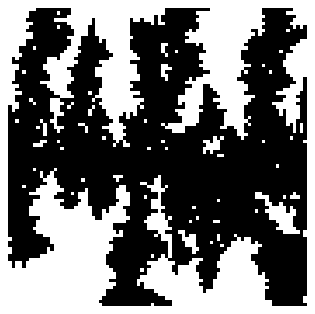}
  
  \hspace*{-15mm}
  \subfe{Fig2d}{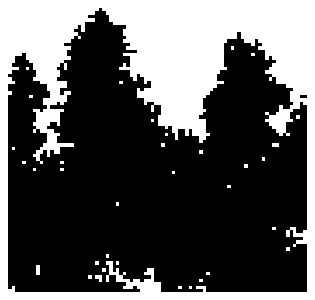}
  \subfe{Fig2e}{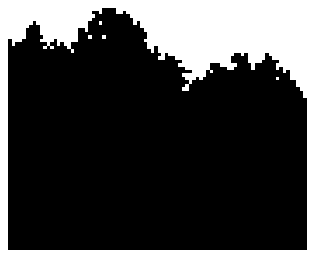}
  \subfe{Fig2f}{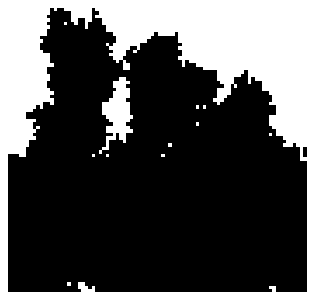}
  \subfe{Fig2h}{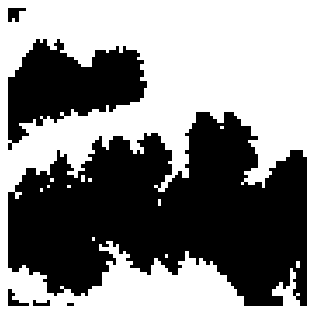}
  \subfe{Fig2k}{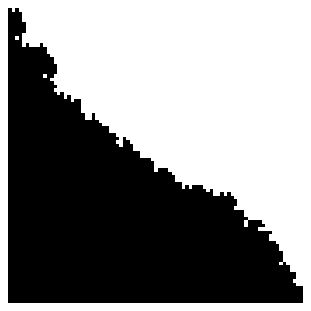}

  \hspace*{-15mm}
  \subfe{Fig3a}{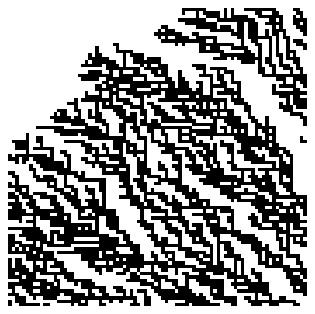}
  \subfe{Fig3b}{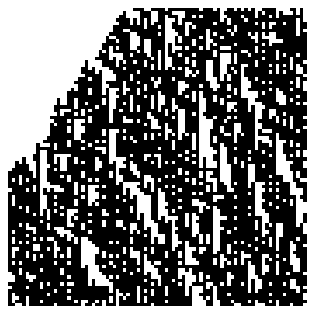}
  \subfe{Fig3c}{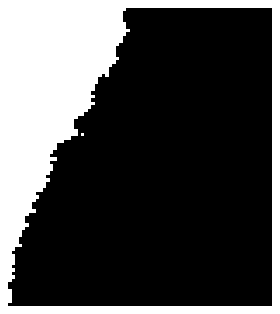}
  \subfe{Fig4a}{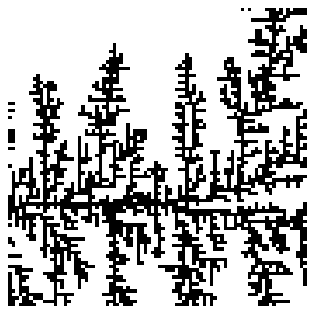}
  \subfe{Fig4c}{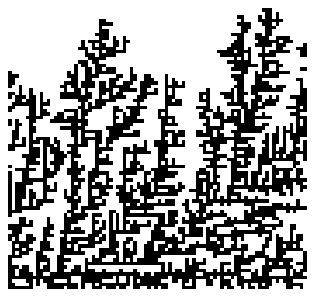}

  \hspace*{-15mm}
  \subfe{Fig4e}{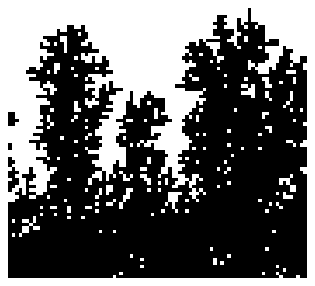}
  \subfe{Fig4f}{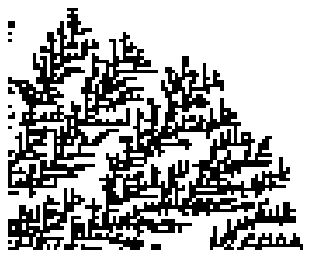}
  \subfe{Fig4i}{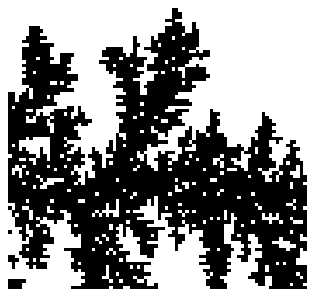}
  \subfe{Fig5a}{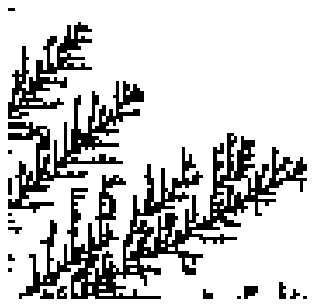}
  \subfe{Fig5c}{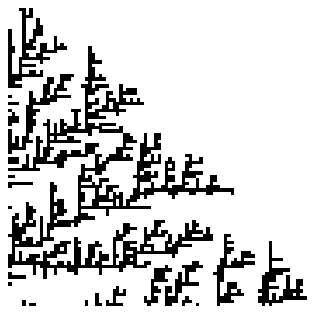}
 
 \hspace*{-15mm}
  \subfe{Fig5d}{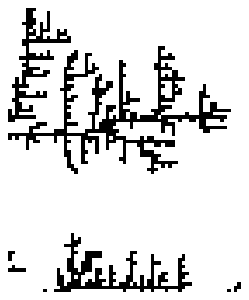}
  \subfe{Fig5g}{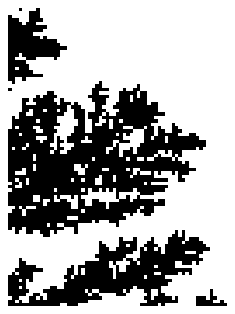}
  \subfe{Fig6b}{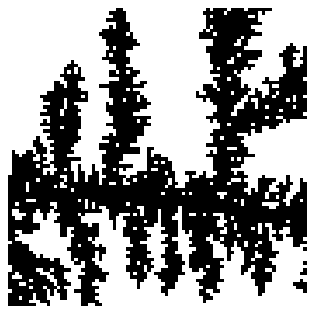}
  \subfe{Fig6c}{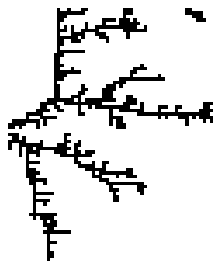}
  \subfe{Fig6d}{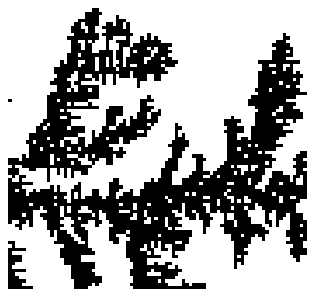}
  \caption{ Enlargements of several clusters from previous figures (with the
    relevant references).}
\end{figure}

\section{A classification of DLA-type growth models}
\label{sec:Classification-DLA-ty}

The standard growth models (DLA and the Eden model) have been
generalised using a variety of ideas.  For example, the Eden model was
modified in \cite{jia:gan:ben89} to obtain results which are similar
to realisations of the DLA.  This was achieved by amending
probabilities of attaching neighbouring sites according to their
positions in the complement of the cluster, more exactly, by taking
into account the number of different paths that lead outside from any
given point.  The corresponding model was called the screened Eden
model.

It is possible to assume that a particle reaching the cluster sticks
to it with probability $p$ and otherwise continues walking so that
every time it touches the cluster again it has an option to stick with
probability $p$. This model is called a penetrable DLA, and yields
somewhat thicker clusters similar to the model with jumps suggested
here.  It is also possible to give particles an option to leave the
cluster with a certain probability. This leads to DLA with
disaggregation.
 
A magnetic DLA model was introduced in \cite{van:aus95} by assuming
that particles may have two different spins and at each stage when new
particles stick to the configuration their spins are chosen at random
according to the corresponding Ising distribution.

It should be noted that all these modifications concern three basic
components of the model: the type of stochastic process that
determines movements of the particle, the lattice structure and the
nature of aggregation. By coincidence, these three ingredients appear
in the same order in the abbreviation DLA, the first letter meaning
diffusion, the second determining the lattice limitation, and the
third concerning aggregation. This leads to a classification of
DLA-type growth models.  For applied probabilists the classification
below reminds us of D.G.~Kendall's famous classification of queueing
systems, see, e.g., \cite{bun96}.

\paragraph{Stochastic process.}
Although a general stochastic process can be used to describe
particles' movements, in many important cases the process has three
possible components: diffusion, jumps and flights. The diffusion
component is determined by the transfer vector and the diffusion
matrix, the jumping component is determined by the distribution of the 
jump (both the direction and the length), and the flight (or ballistic)
component is determined by the direction and the length of the
flight. 

If the diffusion component is the standard Brownian motion, then it is
denoted by ${\mathsf{D}}$. For a general diffusion process one writes
${\mathsf{D}}({\mathbf a},{\mathbf b})$, where ${\mathbf a}$ is the
transfer vector and ${\mathbf b}$ is the diffusion matrix.

The jump component is denoted by ${\mathsf J}$ with indications in
parenthesis of the distribution of the jump. The isotropic
distribution of jumps is indicated by letter $i$ with the rest
determining the distribution of the length of the jump. For example,
${\mathsf J}(i,\Un(0,10))$ indicates that jumps have isotropic
directions and their lengths are uniformly distributed on $(0,10)$. In
a more general case, one writes ${\mathsf J}(i,F)$, where $F$ is the
probability distribution of the length of the jump. Furthermore,
${\mathsf J}(F)$ stands for a general (possibly anisotropic)
distribution with $F$ specifying the distribution of the jump vector.

The flight component is denoted by ${\mathsf B}$ with the same
structure of the terms in parentheses as for the jump component. In
addition, the letter $c$ stands for the centre-bound flights (or jumps).

The three components described above can be combined to produce a
process which may have paths of different kinds. For example,
${\mathsf D}{\mathsf J}(i,10){\mathsf B}(c,20)[0.3,0.4,0.1]$ indicates
that at every step the process moves as a (standard Brownian)
diffusion with probability $0.3$, jumps with probability $0.4$ and
flies towards the centre with probability $0.1$. In the continuous
case we no longer have integer time moments, so these probabilities
give rise to Poisson processes on $[0,\infty)$ whose points determine
time moments of jumps, flights and diffusion movements. It is also
possible to have the fourth component corresponding to the white noise
leading to the Eden model. This component is denoted by ${\mathsf
  W}$. 

On the other hand, rather than switching different trajectories while
moving, we can assume that every new particle adheres to only one type
of its path which is chosen randomly when the particle is launched.
This is denoted by ${\mathsf D}[0.3]{\mathsf J}(i,10)[0.4]{\mathsf
  B}(i,20)[0.1]$ and means that every new particle with probability
$0.3$ undertakes Brownian motion (until it sticks to the cluster
or leaves the active area), with probability $0.4$ moves in jumps of
size 10 only and with probability $0.1$ flies isotropically at length
20 at every step.

If movements of any kind are discarded for particles that do not hit
the cluster while moving (as in the shooter model), then the
corresponding letter is marked with a zero subscript, so that the
shooter model with $p_{\mathrm{b}}=0.1$ and no jumps is denoted by
${\mathsf D}{\mathsf B}_0[0.9,0.1]$.

\paragraph{Lattice structure.}
To define the growth model it is necessary to specify the lattice
where particles move and also to describe the connectivity
relationship.  Discrete lattice models are denoted by ${\mathsf L}$
with a subscript specifying the type of lattice: square (${\mathsf
  S}$), triangular (${\mathsf T}$) or hexagonal (${\mathsf H}$). It is
also possible to use different neighbourhood structures for the
pixels.  For example, on the square lattice it is possible to regard
two pixels as neighbours if they share a common edge or just a common
vertex. In the first case (on the planar grid) every pixel has 4
neighbours (${\mathsf L}_{\mathsf S}[4]$), while in the other case
every pixel has 8 neighbours (${\mathsf L}_{\mathsf S}[8]$).

For off-lattice models (${\mathsf O}$) one has to specify the size $r$
of the ball that is associated with the particles and also the size
$R$ at which a moving particle becomes a neighbour of the cluster. In
order to get a connected cluster, $R$ should be smaller than or equal
to $r$. The relevant notation is ${\mathsf O}_r[R]$.

Note that all models considered in Sections~~\ref{sec:DLA-with-jumps}
and~\ref{sec:general-model-with} can be denoted by ${\mathsf
  L}_{\mathsf S}[4]$ with respect to their lattice structure.

\paragraph{Aggregation rules.}
A number of models considered in the literature adhere to several
basic aggregation rules. The first possibility is that the particle
attaches irreversibly to the cluster if it becomes its neighbour. This
is denoted by ${\mathsf A}_1$. For the noise reduced models we write
${\mathsf A_1}[n]$, where $n$ is the minimal number of particles
accumulated at a given pixel to have this pixel attached to the
cluster.  This is the case for all models considered in previous
sections, which can be denoted by ${\mathsf A}_1[4]$.  Furthermore,
rule ${\mathsf A}_p$ means that a particle becomes a member of the
cluster with probability $p$ and with probability $1-p$ continues
travelling (possibly through the cluster) so that at any time when it
is in neighbouring position to the cluster it can again become
attached to it with probability $p$ independently of previous
attempts.

It is also possible to use random rules for determining aggregation of
pixels. For example, if we define an energy of a configuration of
points, then a new neighbour will be attached to the configuration if
it reduces the cluster's energy. Otherwise the particle continues
travelling. This rule is denoted by ${\mathsf E}$.

%Possible disaggregation of particles is denoted by ${\mathsf G}_q$
%where $q$ is a (small) probability of disaggregation, when a particle
%already present in the cluster becomes detached from it and moves
%according to the underlying stochastic process.

\paragraph{Combinations and extensions.}
The three ingredients described above should be combined to give a
complete description of the model. Then the standard DLA on a square
grid without noise reduction is designated by ${\mathsf D}/{\mathsf
  L_{\mathsf S}}[4]/{\mathsf A}_1$, the Eden model becomes ${\mathsf
  W}/{\mathsf L_{\mathsf S}}[4]/{\mathsf A}_1$, and the ballistic
aggregation is ${\mathsf B}(c,\infty)/{\mathsf L_{\mathsf
    S}}[4]/{\mathsf A}_1$.

A more complicated construct such as ${\mathsf D}{\mathsf
  J}(i,10){\mathsf B}_0(c,20)[0.5,0.4,0.1]/{\mathsf L_{\mathsf
    H}}[6]/{\mathsf A}_{0.6}[5]$ designates a shooting model where at
every step the particle is doing a nearest neighbour walk on the
hexagonal lattice with probability 0.5, jumps in a uniformly
distributed direction a distance 10 with probability 0.4 and with
probability 0.1 flies towards the centre with probability 0.1, so that
the flights are discarded if the particle does not hit the cluster.
The noise reduction parameter is 5 and new particles stick to the
cluster with probability 0.6.

\section{Concluding remarks}
This paper presents just a first exploration of the general model of
growth which can lead to patterns with possibly fractal structure. The
model has a number of free parameters which calls for extensive
simulations. Further investigation is required to figure out fractal
properties of the model in relation to the model parameters and also
to work out statistical techniques suitable for estimating parameters
of the model. 

\subsection*{Acknowledgement}

The author is grateful to D.M. Titterington and J.N.~Chapman for
helpful comments on the manuscript.

%\bibliography{/home/ilya/tex/bibdata/abbrev,/home/ilya/tex/bibdata/x/gesamt}
\newcommand{\noopsort}[1]{} \newcommand{\printfirst}[2]{#1}
  \newcommand{\singleletter}[1]{#1} \newcommand{\switchargs}[2]{#2#1}

\end{document}